\documentclass[aps,pra,twocolumn,showpacs,amsmath]{revtex4}
\usepackage{graphicx}
\begin{document}
\setlength{\arraycolsep}{2pt}
\title{Quantum Teleportation of Light}
\author{Changsuk Noh$^{\dagger}$}
\author{A. Chia$^{\S}$}
\author{Hyunchul Nha$^{\ddagger}$}
\author{M. J. Collett$^{\dagger}$}
\author{H. J. Carmichael$^{\dagger}$}
\affiliation{$^{\dagger}$Department of Physics, University of Auckland,
Private Bag 92019, Auckland, New Zealand\\
$^{\S}$Centre for Quantum Dynamics, Griffith University, Nathan, Queensland 4111, Australia\\
$^{\ddagger}$Department of Physics, Texas A \& M University at Qatar, Dohar, Qatar} 
\date{\today}

\begin{abstract}
Requirements for the successful teleportation of a beam of light, including its
temporal correlations, are discussed. Explicit expressions for the degrees of
first- and second-order optical coherence  are derived. Teleportation of an
antibunched photon stream illustrates our results.
\end{abstract}
\maketitle
\narrowtext

Although proposed originally by Vaidman \cite{Vaidman94} as
a protocol for teleporting the wavefunction of a particle,
continuous-variable teleportation has primarily been developed
as a scheme for teleporting the quantum state of light. Braunstein
and Kimble \cite{Braunstein98} recognized how Vaidman's idea
might be implemented to teleport the state of a single mode
of the electromagnetic field employing squeezed-state entanglement.
Subsequent experiments have largely followed their single-mode
prescription \cite{Furusawa98,Bowen03,Furusawa07}, although these
experiments are inherently broadband and suggest a generalization
from the teleportation of quantum {\it states\/} to the
teleportation of quantum {\it fields\/}.

A broadband theory was developed by van Loock {et al.}
\cite{vanLoock00} and  teleportation of broadband entanglement
has recently been reported \cite{Yonezawa07}. Neither of these
works, however, fully explores the implications of the
generalization to quantum fields. It is our aim in this
Letter to develop that generalization. Specifically, we propose
teleporting a quantum field (beam of light) in the following sense:
as a consequence of Alice's measurements and Bob's actions,
a light beam emerges at the teleporter output carrying over all
the statistical properties of the beam at the input; in this
the quantum field is characterized by correlation functions
in time---ideally to all orders---rather than by a single-mode
quantum state.

Naively one might expect this to be achieved so long as the
squeezing bandwidth is larger than the bandwidth of the input
light. This, however, is not the case, as vacuum modes are
equally a part of a quantum field; for the purposes of
teleportation as proposed, no quantum field may be viewed
as having a bandwidth less than the squeezing. What is required,
then, is optical filtering by Bob: we propose that the
vacuum lying outside the input light bandwidth not be teleported,
but rather be replaced by an equivalent vacuum reflected by
Bob's filter.

Experiments to date take a much different view. They (as Victor
\cite{Furusawa98}) measure properties of the teleported light
over a narrow span of frequencies only, i.e., they filter
in detection. Also, they measure only quadrature phase amplitudes.
Thus, they leave aside temporal correlations
and, in particular, correlations to be described by operators
in different orders. Our aim is to expand upon their point
of view: we ask that {\it any\/} conceivable measurement at
the input yield equivalent results at the output; then one
can say the light (quantum field) is teleported.

{\it Any\/} conceivable measurement is a tall order when the
infinite dimensions of a field are considered. In this
communication we focus on photon correlations, specifically a
measurement of $g^{(2)}(\tau)$. We ask whether and under what
conditions an antibunched photon stream at the input will appear,
with high fidelity, as an antibunched photon stream at the output.
The anticorrelation of photon arrival times places high demands
on the reproduction of the input quantum field across its entire
bandwidth, including all vacuum modes. Moreover, as an expression
of the particle aspect of light, photon correlations play as
counterpoint to the wave amplitude measurements upon which
continuous variable teleportation and its characterization in
experiments to date are based; they require attention
be given to the issue of operator order.

\begin{figure}[t]
\includegraphics*[width=3.2in,keepaspectratio=true]{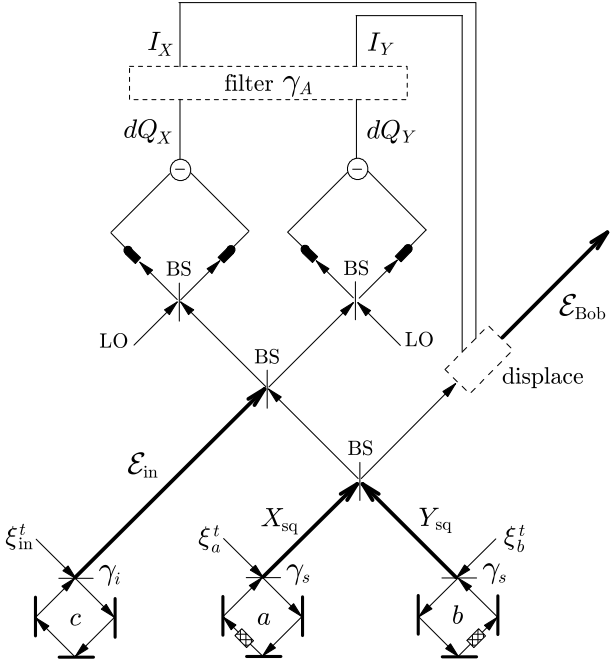}
\caption{Schematic of continuous variable teleportation with no filtering
of the output by Bob; BS: $50/50$ beam splitter; LO: local oscillator.}
\label{fig:fig1}
\end{figure}

We begin by outlining our model of the standard continuous
variable teleporter \cite{Furusawa98}. We work in the time
domain and for the purpose of setting notation, start with all
fields, including vacuum fields, represented by classical stochastic
processes. These are Wigner stochastic processes in conventional
terms, where, for the time being, positive-definite Wigner functions
are assumed; they also conform to the viewpoint of stochastic
electrodynamics \cite{Carmichael03}. A sketch of the model appears
in Fig.~\ref{fig:fig1}. In units of photon flux, input and squeezed
fields are written as
\begin{eqnarray}
{\cal E}_{\rm in}&=&\sqrt{2\gamma_i}c-\xi_{\rm in}^t,\label{eqn:input_fields_a}
\nonumber\\
\noalign{\vskip2pt}
X_{\rm sq}&=&\sqrt{2\gamma_s}a-\xi_a^t,\\
\noalign{\vskip2pt}
Y_{\rm sq}&=&\sqrt{2\gamma_s}b-\xi_b^t\nonumber\label{eqn:input_fields_c},
\label{eqn:input_fields}
\end{eqnarray}
where $2\gamma_i$ and $2\gamma_s$ are field bandwidths, and $\xi_{\rm in}^t$,
$\xi_a^t$, $\xi_b^t$ are Gaussian while noises, i.e., vacuum fields.
Alice measures the complex valued photocurrent $I=I_X+iI_Y$,
with
\begin{equation}
dI/dt=-\gamma_A(I-dQ/dt),
\label{eqn:current_filter}
\end{equation}
where $2\gamma_A$ is her measurement bandwidth, and
\begin{equation}
\frac{dQ}{dt}=\frac1{\sqrt2}{\cal E}_{\rm in}+\frac12(X_{\rm sq}+Y_{\rm sq})^*.
\label{eqn:current_increment}
\end{equation}
Bob implements the displacement $\sqrt2 I$ to produce
\begin{eqnarray}
{\cal E}_{\rm Bob}&=&\frac1{\sqrt2}(X_{\rm sq}-Y_{\rm sq})\nonumber\\
&&+F_A*\left[{\cal E}_{\rm in}+\frac1{\sqrt2}(X_{\rm sq}+Y_{\rm sq})^*\right],
\label{eqn:EBob}
\end{eqnarray}
where $F_A(t)$ is the impulse response of Alice's filter and $*$ denotes
convolution.  If we then assume a broad measurement bandwidth for Alice,
$\gamma_A\gg\gamma_s,\gamma_i$, approximating $F_A(t)$ as a $\delta$-function
yields
\begin{equation}
{\cal E}_{\rm Bob}={\cal E}_{\rm in}+\sqrt2\left(X_{\rm sq}^X-iY_{\rm sq}^Y\right).
\end{equation}

It seems that perfect squeezing, $X_{\rm sq}^X=Y_{\rm sq}^Y=0$,
leads to perfect teleportation,  ${\cal E}_{\rm Bob}={\cal E}_{\rm in}$.
We must note the assumption $\gamma_A\gg\gamma_s$, however, since according to
it the squeezing cannot be perfect across Alice's entire bandwidth.
Nevertheless, the ideal result can  be recovered if $\gamma_s=\gamma_A$ and
both Alice's spectrum and the squeezing spectrum are flat.
Then, for perfect squeezing,
\begin{equation}
{\cal E}_{\rm Bob}=\left\{
\begin{matrix}
{\cal E}_{\rm in},&\qquad|\omega|<\gamma_s=\gamma_A\\
\noalign{\vskip2pt}
(\xi_a^t-\xi_b^t)/\sqrt2,&\qquad|\omega|>\gamma_s=\gamma_A,
\end{matrix}
\right.
\label{eqn:matched_result}
\end{equation}
where here and in what follows $\omega$ is measured relative to the input
light line center. Outside the squeezing bandwidth, Eq.~(\ref{eqn:matched_result})
simply replaces $X_{\rm sq}$ and $Y_{\rm sq}$ [first term on the right in
Eq.~(\ref{eqn:EBob})] by the vacuum fields $\xi_a^t$ and $\xi_b^t$. The
result, being a vacuum field, is operationally equivalent to teleporting
${\xi}_{\rm in}^t$.

\begin{figure}[b]
\includegraphics*[width=2.4in,keepaspectratio=true]{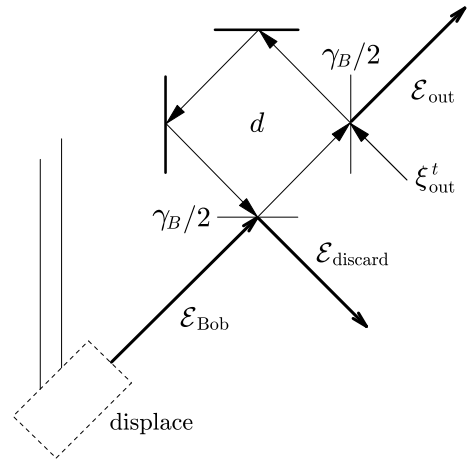}
\caption{Schematic of Bob's filter.}
\label{fig:fig2}
\end{figure}

This matched-spectra scenario is highly idealized, though, and it is more
realistic to take $\gamma_A>\gamma_s$ and allow both spectra a frequency
roll off; Alice's bandwidth must of course exceed the squeezing bandwidth,
otherwise it is not possible to completely cancel the squeezed-light
noise distributed to Bob. Under these conditions, Bob might then filter his
field (Fig.~\ref{fig:fig2}), to generate the output
\begin{equation}
{\cal E}_{\rm out}=F_B*\left({\cal E}_{\rm Bob}+\xi_{\rm out}^t\right)
-\xi_{\rm out}^t,
\label{eqn:Eout}
\end{equation}
with $F_B(t)$ the impulse response of Bob's filter; the filter bandwidth is
$2\gamma_B$ and $\xi_{\rm out}^t$ is the vacuum reflected at its
output. 

We now ask under what conditions does ${\cal E}_{\rm out}={\cal E}_{\rm in}$
or ${\cal E}_{\rm out}=\xi_{\rm out}^t$, $|\omega|\gg\gamma_i$, and
does such perfect teleportation carry over to the teleportation of an
antibunched photon stream? To be more explicit, we might set ${\cal E}_{\rm in}$ 
as the scattered field from a driven two-state atom (resonance fluorescence).
What then are the requirements on the employed squeezing and relative bandwidths 
such that Bob's filter (mode $d$ of Fig.~\ref{fig:fig2}) behaves as a two-state
system, i.e., is never occupied by more than one photon?

In addressing these questions the classical (Wigner) representation of
stochastic processes must be dropped. Strictly, only $I$ and $Q$ in the
above equations are c-numbers; all the fields are operators. If, then,
one aims to {\it realize\/} Alice's classical information, an explicit
modeling of her measurements is needed to connect the operators to the
c-numbers. Quantum trajectory theory does this with 
\begin{equation}
\frac{dQ}{dt}=\frac1{\sqrt2}\langle\hat{\cal E}_{\rm in}\rangle_{\rm rec}
+\frac12(\langle\hat X_{\rm sq}^\dagger\rangle_{\rm rec}
+\langle\hat Y_{\rm sq}^\dagger\rangle_{\rm rec})-\xi_{\rm shot}^t
\label{eqn:trajectory_current}
\end{equation}
in place of Eq.~(\ref{eqn:current_increment}). Note the shot noise,
$\xi_{\rm shot}^t$, in place of the classical representation of vacuum
fluctuations, $\xi_{\rm in}^t/\sqrt2+(\xi_a^{t*}+\xi_b^{t*})/2$. The
difficulty with this approach is that although low fidelity teleportation
can be simulated \cite{Carmichael05}, numerical simulation of high-fidelity
conditions is impractical. In response we adopt an alternative method
of calculation that while it does not realize Alice's classical information,
does provide analytical expressions for correlation functions---for
any operator order. We replace the classical fields ${\cal E}_{\rm in}$,
$X_{\rm sq}$, $Y_{\rm sq}$, etc. by operators, and the c-numbers $I$ and
$Q$ by formal operators \cite{Note1}. Then the same linear map from
input to output holds as an operator expression. Output field correlation
functions may be determined from the known correlation functions of
$\hat{\cal E}_{\rm in}$, $\hat X_{\rm sq}$, and $\hat Y_{\rm sq}$.

\begin{figure}[b]
\includegraphics*[width=1.6in,keepaspectratio=true]{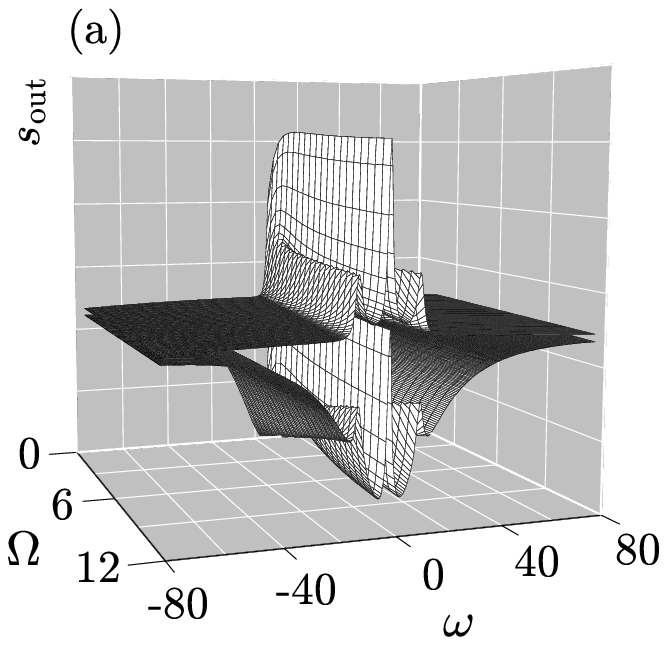}\hskip0.2in
\includegraphics*[width=1.6in,keepaspectratio=true]{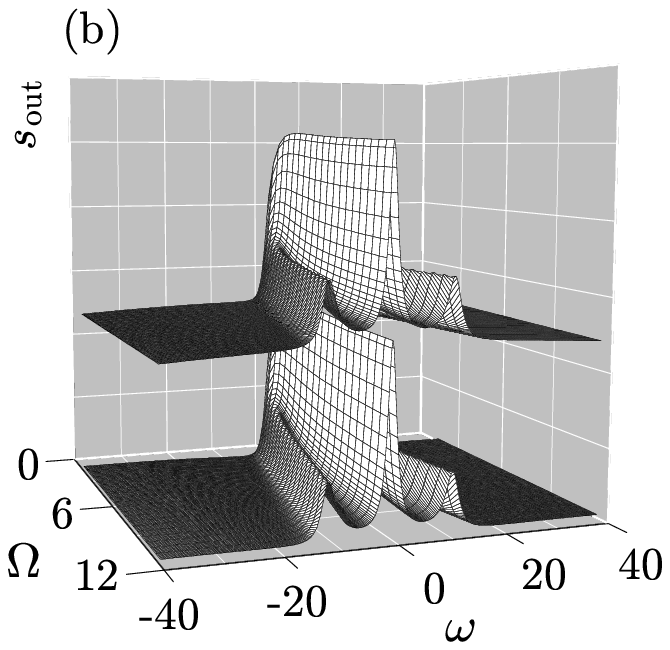}
\vskip0.2in
\includegraphics*[width=1.6in,keepaspectratio=true]{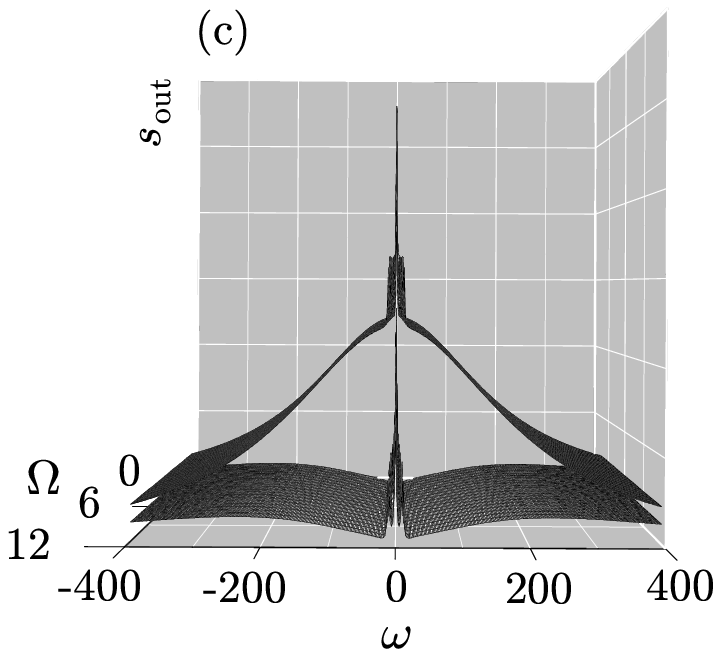}\hskip0.2in
\includegraphics*[width=1.6in,keepaspectratio=true]{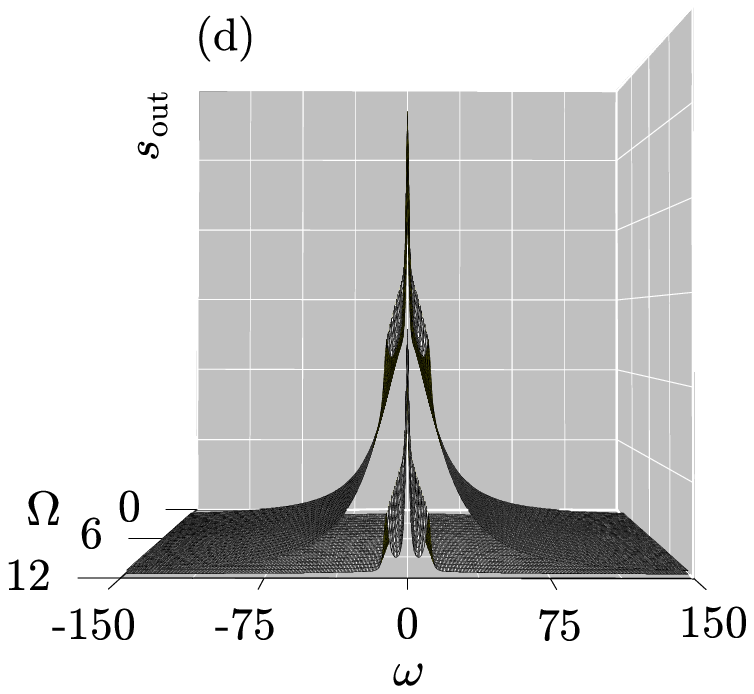}
\caption{Teleportation of the Mollow spectrum: $s_{\rm out}(\omega)$ is plotted
for $(\gamma_i,\gamma_A,\gamma_B)/\gamma_s=(0.1,50,100)$ (a); $(0.005,5,1)$ (b,c);
and $(0.005,5,0.1)$ (d). Upper and lower surfaces correspond, respectively,
to $0\mkern2mu{\rm dB}$ and $25\mkern2mu{\rm dB}$ line-center squeezing.
The Rabi frequency and frequency, $\Omega$ and $\omega$, are in units of $\gamma_i$.}
\label{fig:fig3}
\end{figure}

\begin{figure}[t]
\includegraphics*[width=1.6in,keepaspectratio=true]{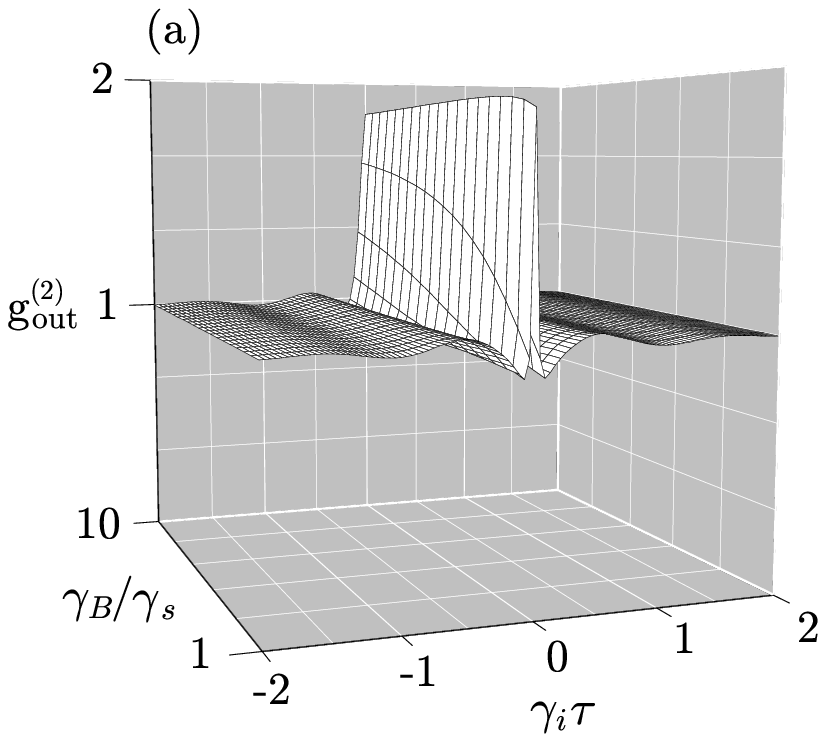}\hskip0.2in
\includegraphics*[width=1.6in,keepaspectratio=true]{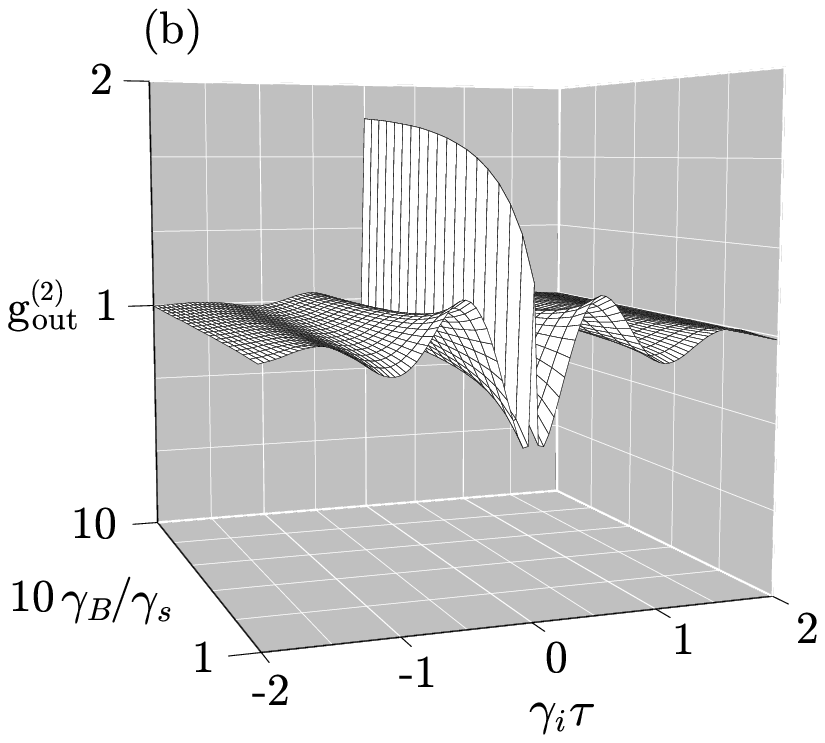}
\vskip0.2in
\includegraphics*[width=1.6in,keepaspectratio=true]{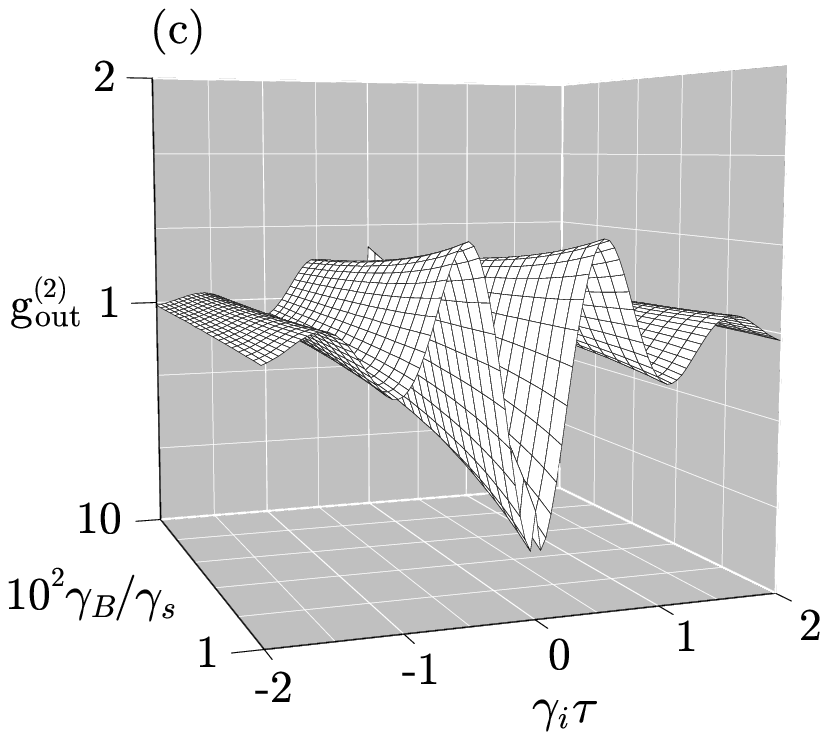}\hskip0.2in
\includegraphics*[width=1.6in,keepaspectratio=true]{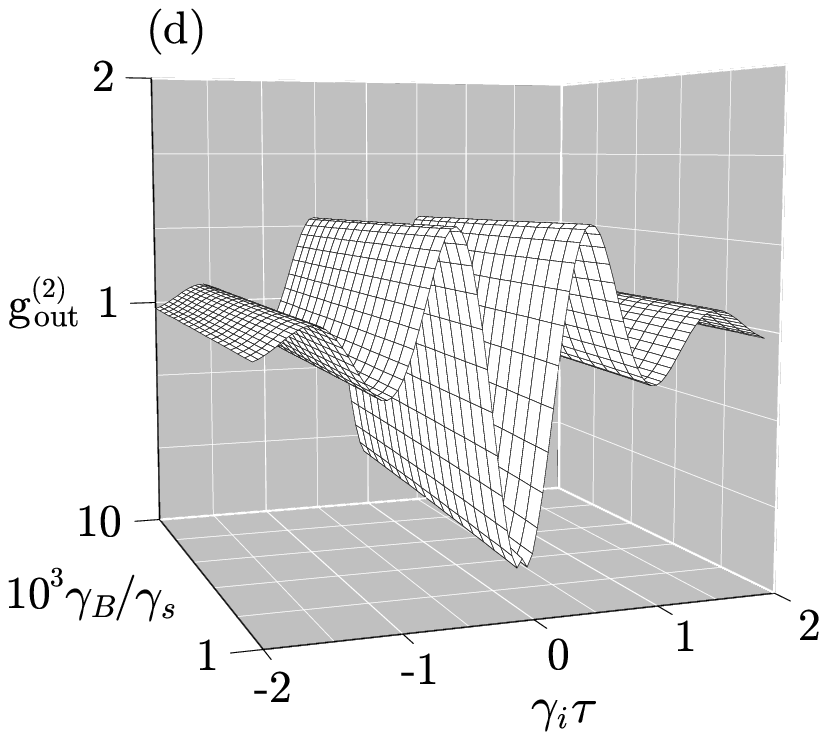}
\caption{Teleportation of photon antibunching:
$g^{(2)}_{\rm out}(\tau)$ is plotted for $25\mkern2mu{\rm dB}$ line-center
squeezing, $\gamma_A/\gamma_s=5$, $\gamma_B/\gamma_i=20$, and $\gamma_B/\gamma_s$
varying over four orders of magnitude [(a)-(d)]; Rabi frequency
$\Omega/\gamma_i=6$.}
\label{fig:fig4}
\end{figure}

We have carried this program through to derive general expressions,
functions of all parameters, for the degrees of first and second-order
optical coherence. While these expressions are used to generate the plots
of Figs.~\ref{fig:fig3} and \ref{fig:fig4}, they are too unwieldy to
reproduce in general form. We present them simplified for large
relative bandwidths, $\gamma_A\gg\gamma_B,\gamma_s,\gamma_i$, and
$\gamma_B\gg\gamma_i$. Then if $g^{(1)}_{\rm in}(\tau)$ and
$g^{(2)}_{\rm in}(\tau)$ are the input field degrees of first- and
second-order optical coherence, we find for the teleported light:
\begin{equation}
f_{\rm out}g^{(1)}_{\rm out}(\tau)=f_sA_s(\tau)+f_{\rm in}g^{(1)}_{\rm in}(\tau),
\label{eqn:first-order}
\end{equation}
where $f_{\rm in}=\langle\hat{\cal E}_{\rm in}^\dagger\hat{\cal E}_{\rm in}\rangle$
and $f_{\rm out}=\langle\hat{\cal E}_{\rm out}^\dagger\hat{\cal E}_{\rm out}\rangle$
are input and output field photon fluxes, and
\begin{equation}
\frac{f_s}{\gamma_B}A_s(\tau)=\frac{\frac12(\gamma_B^2-\gamma_-^2)e^{-\gamma_B\tau}
-\frac{2\lambda}{1+\lambda}\gamma_B\gamma_se^{-\gamma_+\tau}}{\gamma_B^2-\gamma_+^2},
\label{eqn:Asubs}
\end{equation}
with $A_s(0)\equiv1$ and $\gamma_{\pm}\equiv\gamma_s(1\pm\lambda)$; and 
\begin{eqnarray}
g^{(2)}_{\rm out}(\tau)&=&1+\left(\frac{f_{\rm in}}{f_{\rm out}}\right)^2
[g^{(2)}_{\rm in}(\tau)-1]\nonumber\\
&&+\frac{f_s}{f_{\rm out}}A_s(\tau)\left[g^{(1)}_{\rm out}(\tau)+\frac{f_{\rm in}}
{f_{\rm out}}g_{\rm in}^{(1)}(\tau)\right].
\label{eqn:second-order}
\end{eqnarray}
The parameter $\lambda$ determines the degree of line-center squeezing
at the output of a sub-threshold parametric oscillator, which is given
in dB by $-20\log\left[(1-\lambda)/(1+\lambda)\right]$. 

The spectrum of the teleporter output light is the Fourier transform
of $g^{(1)}_{\rm out}(\tau)$. We discuss it first to illustrate the interplay
of the various bandwidths, $\gamma_A$, $\gamma_B$, $\gamma_i$, and $\gamma_s$.
We consider the example of input resonance fluorescence, i.e., teleportation of
the Mollow triplet. Results are displayed in Fig.~\ref{fig:fig3}, where the
incoherent part of the spectrum, $s_{\rm out}(\omega)$, is plotted as a function
of Rabi frequency $\Omega$. Quite generally, from Eq.~(\ref{eqn:first-order}),
the output spectrum is the spectrum of the input light sitting on a noise background.
The relative size of the pieces (integrated over $\omega$) is
\begin{equation}
\frac{f_s}{f_{\rm in}}=\frac{\gamma_B}{\eta\gamma_i}\left(\frac12-\frac{2\lambda}
{1+\lambda}\frac1{1+\lambda+\gamma_B/\gamma_s}\right)\frac{\Omega^2+2\gamma_i^2}
{\Omega^2},
\label{eqn:fsfinratio}
\end{equation}
from Eq.~(\ref{eqn:Asubs}) and the photon flux directed to the teleporter input,
$f_{\rm in}=\eta\gamma_i \Omega/(\Omega^2+2\gamma_i^2)$; $2\gamma_i$ is the
Einstein $A$ coefficient and $\eta<1$ is a collection efficiency. For $\lambda=0$
(classical teleportation), the factor $1/2$ inside the bracket is the mean number
of photons, $\langle\hat d^\dagger\hat d\rangle=1/2$, in Bob's filter. It may be
traced, formally, to the vacuum fluctuations presented to Alice---a product of the
conjugate on the dispersed-field contribution, $(\hat X_{\rm sq}^\dagger+
\hat Y_{\rm sq}^\dagger)/\sqrt2$, in the formal operator representation of
Alice's photocurrent [Eqs.~(\ref{eqn:current_increment}) and (\ref{eqn:EBob})];
physically, however, it is a number of real photons generated through 
Bob's displacement by Alice's shot noise---$\xi_{\rm shot}^t$ in
Eq.~(\ref{eqn:trajectory_current}) [no photons travel to Bob via the dispersed
field $(\hat X_{\rm sq}-\hat Y_{\rm sq})/\sqrt2$].

The aim in {\it quantum\/} teleportation  is to remove the background.
Figure~\ref{fig:fig3} illustrate how this might be done. In frame (a), for
any $\lambda>0$, the background is reduced over the squeezing bandwidth.
Here the displacement driven by Alice's photocurrent partially nulls the light
now dispersed to Bob; but a residual background persists across Alice's detector
bandwidth (with $\gamma_B>\gamma_A$). In frames (b) and (c) filtering
($\gamma_B<\gamma_A$) suppresses it near line center, although not fully
in the wings. In frame (d), with $\gamma_A\gg\gamma_s\gg\gamma_B\gg\gamma_i$
and sufficient squeezing,  near perfect teleportation of the Mollow triplet
is achieved; note, for example, that the bracket in Eq.~(\ref{eqn:fsfinratio})
goes to zero for $\gamma_B/\gamma_s\to0$ and $\lambda\to1$.

Turning now to the teleportation of an antibunched photon stream,
$f_s/f_{\rm in}$ is again the figure of merit: as it approaches zero,
$g_{\rm out}^{(2)}(\tau)\to g_{\rm in}^{(2)}(\tau)$ [Eqs.~(\ref{eqn:first-order})
and (\ref{eqn:second-order})]. What, however, are quantitatively
the conditions under which Bob's filter will act as a two-state system, i.e.,
exhibit near perfect short-time photon anticorrelation? Short times map to the spectral
wings, where unfiltered light from Alice's shot noise is 
highly detrimental. The front curves of frames (a) and (b)
in Fig.~\ref{fig:fig4} correspond to the parameters of frames (c)
and (d) in Fig.~\ref{fig:fig3}. They show a prominent spike at zero delay,
i.e., bunched rather than antibunched photons. Stronger filtering
is needed to teleport photon antibunching [Figs.~\ref{fig:fig4}(c) and (d)],
though, even then, at $20\mkern2mu{\rm dB}$ line-center squeezing
$g^{(2)}_{\rm out}(0)=0$ is not recovered.

Requirements to achieve $g^{(2)}_{\rm out}(0)\approx0$  follow from
\begin{equation}
g^{(2)}_{\rm out}(0)=2\left[1-\left(1+f_s/f_{\rm in}\right)^{-2}\right]
\end{equation}
[use $A_s(0)=g_{\rm in}^{(1)}(0)=1$ and  $g^{(2)}_{\rm in}(0)=0$ in
Eqs.~(\ref{eqn:first-order}) and (\ref{eqn:second-order})]
and, from Eq.~(\ref{eqn:fsfinratio}), in the limit $\gamma_B/\gamma_s\to0$, 
\begin{equation}
\frac{f_s}{f_{\rm in}}=\frac{\gamma_B}{2\eta\gamma_i}\left(\frac{1-\lambda}{1+\lambda}
\right)^2\frac{\Omega^2+2\gamma_i^2}{\Omega^2}.
\label{eqn:fsfinratio1}
\end{equation}
Thus, a set value of $g^{(2)}_{\rm out}(0)\ll1$ calls for 
\begin{equation}
-20\log\left(\frac{1-\lambda}{1+\lambda}\right)=-10\log\left(\frac{\Omega^2}{\Omega^2+2\gamma_i^2}
\frac{\eta\gamma_i}{2\gamma_B}g^{(2)}_{\rm out}(0)\right)\nonumber
\end{equation}
dB of line-center squeezing. Then Bob's filter bandwidth is  constrained
by the need to approximate the $\gamma_B/\gamma_s\to0$ limit [expand
the bracket in Eq.~(\ref{eqn:fsfinratio})]:
\begin{equation}
\gamma_B/\gamma_s<\frac{2\Omega^2}{\Omega^2+2\gamma_i^2}\frac{\eta\gamma_i}
{2\gamma_B}g^{(2)}_{\rm out}(0).
\end{equation}

\break

\begin{figure}[t]
\vskip0.15in
\includegraphics*[width=3in,keepaspectratio=true]{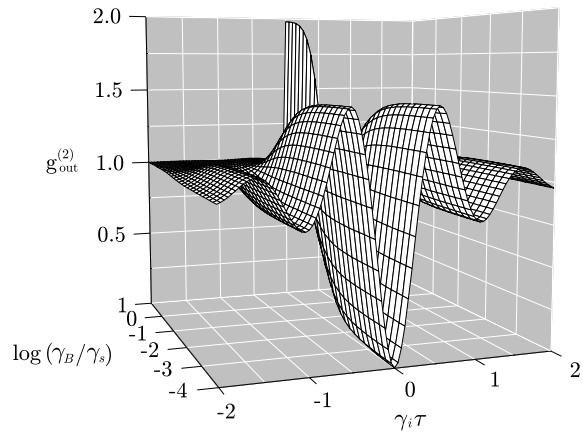}
\caption{As in Fig.~\ref{fig:fig4}; for $46\mkern2mu{\rm dB}$ line-center
squeezing in the limit $\gamma_A/\gamma_s\gg1$, $\gamma_B/\gamma_i\gg1$
[Eq.~(\ref{eqn:second-order})].}
\label{fig:fig5}
\end{figure}

Figure 5 illustrates the recovery of near perfect photon anticorrelation,
with $g^{(2)}(0)\approx0.003$ for $\gamma_B/\gamma_s=10^{-4}$. It
demonstrates the central importance of Bob's filter for teleporting
an antibunched photon stream.  Without it the output has thermal
statistics, $g^{(2)}(0)=2$, even though the Mollow triplet is almost
perfectly reproduced around the center of the spectrum. The spike at
zero delay is caused by a flux of real photons generated by Alice's
shot noise: while Wigner noises might formally represent either the
vacuum fluctuations of a quantized field or photocurrent shot noise
in calculations of quadrature amplitude variances, the two are, in fact,
physically distinct---shot noise but not vacuum noise can produce
the spike in $g^{(2)}(\tau)$ that Bob's filter must remove.

\section*{Acknowledgements}
This work is supported by the Marsden Fund of the RSNZ and HN is supported
by grant NPRP 1-7-7-6 of the Qatar National Research Fund.

\end{document}